# Substitutional Al solute interaction with edge and screw dislocation in Ni: a comparison between atomistic computation and continuum elastic theory

S. Patinet[1]

[1]Service de Recherches de Métallurgie Physique, CEA-Saclay/DEN/DMN
91191 Gif-sur-Yvette Cedex, France, www.cea.fr

## ABSTRACT

Molecular static simulations have been performed to study the interaction between a single dislocation and a substitutional Al solute atom in a pure crystal of Ni. When the Al solute is situated at intermediate distance from the slip plane, we find that both edge and screw dislocations experiment a non-negligible binding energy. We show that for such length scale the description of the elasticity theory can be improved by taking into account the spreading of dislocation cores via the Peierls-Nabarro model.

## 1. Introduction

The mechanical behaviour of metallic alloyed materials is largely caused by the interactions between dislocations and solute atoms. Foreign atoms play the role of obstacles to the dislocations motion and at finite temperatures segregate on dislocation forming Cottrell atmospheres. These atomic processes lead to phenomenon as solid solution hardening, heterogeneous precipitation, static or dynamic ageing and impact the yield stress of the alloy. Therefore, understanding the interaction between solute atoms and dislocation is an important issue in materials science. One of the most recent ways to examine this problem quantitatively is to perform atomistic simulations [1, 2]. Empirical potential based on Embedded Atom Method (EAM) have been found to be reliable to model dislocation in Face Centered Cubic (fcc) metals [1]. Recently, the role played by the screw dislocation segments have been addressed and it has been shown that for a certain model alloy, i.e. Ni(Al) the pinning strength by solutes situated in the vicinity of the core are of same order as for the edge dislocation [2, 3]. On the other hand most of the metal macroscopic behaviors involve collective evolution of dislocations and make atomistic simulations no more tractable. It is therefore interesting to tentatively model the dislocation solute interactions via a continuum elastic theory [4, 5, 6] to check its range of validity and to provide a relevant description of it that may serve at larger scale as for instance into the dislocation dynamics simulations.

## 2. Atomic Simulations

We perform molecular static (MS) relaxations in pure Ni to study the geometry of perfect isolated edge and screw dislocation (see Fig. 1). The calculations are done using EAM potentials supplied by [7] that gives a good description of the Ni(Al) alloy properties. The simulation cell is oriented so as that the horizontal Z planes are the $(1\bar{1}1)$ slip plane of the fcc lattice while the Y direction corresponds to the dislocation line ($a_0/2[110]$, screw dislocation Burgers vector). The X direction is orthogonal to Z and Y and points at the dislocation motion ($a_0/2[110]$, edge dislocation Burgers vector). The simulation box sizes along the directions X, Y, Z are 100×30×50 Å and 80×20×80 Å for the edge and the screw dislocation respectively. Periodic boundary conditions are imposed along X and Y while free surface conditions are applied to the atoms in the upper and lower Z surfaces [1, 2, 3]. The relaxed crystal structure of a single perfect dislocation is shown in Fig. 1 through an analysis of the Burgers vector density of both edge and screw component lying in the slip plane. As expected in Nickel, the full dislocation dissociates into 2 Shockley partials with $a0/6[211]$ and $a0/6[12\bar{1}]$ Burgers vectors separated by a stacking fault region. The core of each partial is far from being compact and spreads in the glide plane over a finite width ξ. The dissociation distance $d$ between partials as well the partial core widths are found larger for the edge dislocation than for the screw.

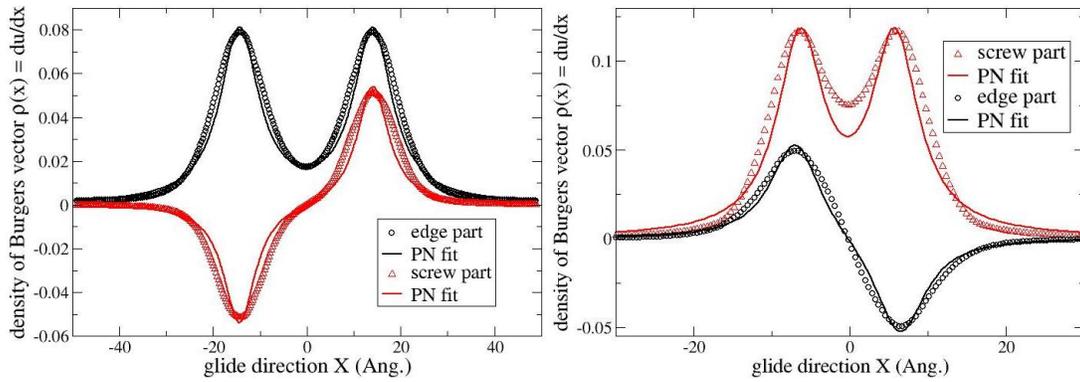

Figure 1. Densities of Burgers vectors with respect to the glide direction in the slip plane: (a) edge dislocation, (b) screw dislocation. The symbols represent the simulation data while the curves are their fits on Peierls-Nabarro displacement field.

In a second time, we carried out the same MS simulations except that a single isolated Al atom is placed in a simulation cell of pure Ni. Then an external applied shear stress is produced by imposing extra forces to the atoms in the upper and lower Z surfaces making the dislocation glide toward the obstacle. At each relaxation step a gradient algorithm forces the dislocation to explore its minimum potential energy path. The interaction energy is recorded for all relative distances between the dislocation and the Al solute atom situated at the third $[1\bar{1}1]$ plane from the glide plane for both dislocations (see Fig. 2). The minimum distance between the solute and the dislocation cores is thus about ~5 Å and corresponds to two and a half of the inter-plane distance along the $[1\bar{1}1]$ direction. As expected, the compressive zones, above the glide plane for the edge dislocation and alternate for each partial of the screw dislocation, are repulsive for the solute while the tensile zones are attractive since Al impurities dilate the Ni matrix. The attractive and repulsive peaks are clearly associated with the positions of the partials. At such distance the maximum interaction energy is found to be -0,046 $eV$ for the edge dislocation while it is twice smaller for the screw. In case of the screw dislocation, the maximum binding energy is greatly increase by the modulus effect. From the results reported in Fig. 2, we can expect the segregation of solutes forming Cottrell atmospheres of different shapes for each dislocation character: below the glide plane for the edge dislocation, above the leading partial and below the trailing one for screw segments. Therefore, the interaction between screw dislocation and solute atoms even at such length scale can not be neglected and the screw segments may as well participate in ageing phenomenon or solid solution hardening.

Table 1. Physical constants calculated from Ni(Al) EAM empirical potential.

| $C_{11}$ GPa | $C_{12}$ GPa | $C_{44}$ GPa | $\dfrac{dC_{11}}{C_{11}dc}$ | $\dfrac{dC_{12}}{C_{12}dc}$ | $\dfrac{dC_{44}}{C_{44}dc}$ | $E$ GPa | $v$ | $\delta V$ Å³ | $d_e$ Å | $d_s$ Å | $\xi_e$ Å | $\xi_s$ Å |
|---|---|---|---|---|---|---|---|---|---|---|---|---|
| 246 | 147 | 125 | -1,87 | -1,31 | -1,28 | 242 | 0,27 | 2,1 | 28,4 | 12,6 | 4,5 | 3,8 |

## 3. Elasticity Theory

The linear continuum elastic theory has shown to be reliable to describe the interaction energy between defects and dislocations as long as deformations are small, i.e. far enough from the dislocation core [4, 5, 6]. Considering at same time size and modulus misfits, the binding energy between a dislocation and a single defect reads as follows [1]:

$$E_b = P\delta V - \frac{1}{2}V\sigma_{ij}\frac{\partial S_{ijkl}}{\partial c}\sigma_{kl}, \tag{1}$$

where P and $\sigma_{ij}$ are the hydrostatic pressure and the stress field created by the dislocation at the solute position, $V$ is the atomic volume in the matrix, $\delta V$ the solute relaxation volume, c the solute concentration and $S_{ijkl}$ is the compliance tensor expressed in this study via isotropic constant (see Tab. 1). Since a substitutional Al atom in fcc Ni does not break the cubic symmetry of the lattice it is not required to account for tetragonal distortion effect. For our isotropic elasticity calculations, we use the Young's modulus $E$ and the Poisson coefficient $v$ obtained by the Voigt average. Taking full account of the periodic boundary conditions as well as the dissociation of dislocation we carried out the computations of types [8]: (i) Volterra dislocation with compact core, (ii) Peierls-Nabarro (PN) dislocation with an extended core region.

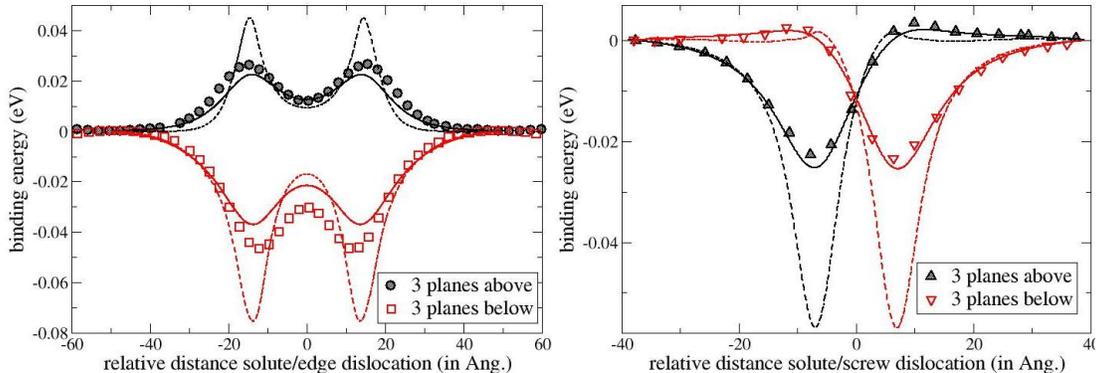

Figure 2. The binding energy between one Al solute atom and a dislocation against the distance between the dislocation and the obstacle in the glide direction: (a) edge dislocation, (b) screw dislocation. The solute is situated in the third ($1\bar{1}1$) planes above (filled symbols) or below (open symbols) the glide plane. The curves correspond to analytical treatments: Volterra dislocation (dashed line) and Peierls-Nabarro dislocation (full line).

In Fig. 2, the predicted elasticity results are found to be in fairly good agreement with the atomistic results. However, several discrepancies need to be pointed out. Due to the singularity of its elastic field, the results given by Volterra dislocation systematically overestimate the interacting energy just above or below the core positions of the partials [5]. On the other hand, we note that the interaction energy dependency described by a PN dislocation gives a better agreement with the simulation data. Indeed, the PN description of a spreaded core is closer from the burgers density profile of the dislocation (see Fig.1). We also remark an overestimation of the interaction energy for the edge dislocation when the solute is situated around the stacking fault region. This seems consistent since the elasticity solutions do not address, in the present study the solute–stacking fault interaction. To improve the transfer from the atomic scale to a continuous theory we emphasize that some improvements of the model would be required as the nonlinear, anisotropic nature of the dislocation-defect interaction as well as the stacking fault elastic field. Despite of simplified assumptions, the PN model has the great merit of providing an analytical nonlinear elastic model of the dislocation core. However at smaller length scales, nearby of the dislocation core, linear elastic theory breaks down and the PN model calculations deviate from our atomistic calculations.

## 4. Conclusions

The aim of the present paper was to extend the comparison between the atomistic computations and elastic theory in the case of a single substitutional solute atom and dislocations in a fcc alloy. For an impurity situated at intermediate distance from the glide plane, we found that the binding energy of the screw dislocation is twice smaller than the edge one. Another issue of the present paper was to tentatively apply analytical models to compute the binding energy. It has been shown that the elasticity theory manages to predict qualitatively the interaction energy and that these predictions can be improved by accounting for the spreading of the dislocation core.